\documentstyle[aps,graphicx,tighten,preprint]{revtex}
\begin{document}
\title{Macroscopic Local Realism Incompatible with Quantum Mechanics: 
Failure of Local Realism where Measurements give Macroscopic 
Uncertainties\\ }
 \vskip 1 truecm
\author{M. D. Reid\\ }

\address{Physics Department, University of Queensland, Brisbane, Australia\\ }  	  
\date{\today}
\maketitle
\vskip 1 truecm
\begin{abstract}
We show that quantum mechanics predicts a contradiction with local 
hidden variable 
theories for photon number measurements which have limited resolving 
power, to the point of 
imposing an uncertainty in the photon number result which is 
macroscopic in absolute terms. We  
show how this can be interpreted as a failure of a new premise, macroscopic 
local realism.
\end{abstract}
\narrowtext
\vskip 0.5 truecm
 Bell $^{\cite{1}}$ in 1966 showed that the premises of local 
realism (or local hidden variable theories) were incompatible with 
the predictions of quantum mechanics.  
Experiments $^{\cite{2}}$ support quantum mechanics, and the general viewpoint is 
to reject the premise of local realism. 

To date theoretical and experimental effort 
has focussed on situations where results of the relevant 
measurements need be only microscopically separated $^{\cite{15}}$. 
The measurements performed are intrinsically 
microscopic, in that one requires to clearly distinguish 
between results (eigenvalues of the appropriate quantum operator) 
which are microscopically distinct.

 Theoretical work has shown a failure of local realism for 
 multi-particle (or higher spin) systems $^{\cite{3}}$, where 
 the system and range of results can be macroscopic. There have also 
 been proposals $^{\cite{4}}$ which show failure of local realism for 
 quantum superpositions of two 
 macroscopically distinct states. However the violations are still 
 apparently 
 only indicated  
 where  
 measurements at some point must resolve microscopically different 
 results, such as adjacent photon number or spin values. 
 While the results indicate failure of 
 local realism for macroscopic systems, it is not clear whether one 
 is testing a premise different to that tested in the microscopic 
 experiments.

Schrodinger $^{\cite{5}}$ raised the issue of quantum mechanics apparently 
predicting superpositions of states macroscopically distinct 
(``Schrodinger-cat states''), questioning the possibility of their 
true existence, based on the notion that such states apparently 
violate a type of macroscopic realism. Recent progress $^{\cite{6}}$ in the 
experimental 
generation of such superpositions highlights a need 
to test objectively for a true 
incompatibility with a macroscopic realism. Progress has been made by 
Leggett and Garg $^{\cite{7}}$ who predict an incompatibility 
of quantum  mechanics 
with the premise of ``macroscopic realism and noninvasive 
measurability''.  

Here we define the premise of ``macroscopic local realism'' in such a way 
that its failure is more surprising than failure of the 
local realism addressed in previous Bell-type studies. This local 
realism becomes immediately testable in experiments where the results of all relevant 
measurements are 
macroscopically distinct, if a failure of local realism in the usual 
way can be shown. 
Experiments which still show a failure of local realism, even 
when uncertainties in all relevant measurements are macroscopic, will also show a 
failure of this type of 
macroscopic local realism. In this paper we prove this result and  
present a quantum state 
with this property, claiming therefore what is to our knowledge 
the first reported predicted failure of such macroscopic local realism.

 In 1935 
 Einstein, Podolsky and Rosen $^{\cite{8}}$ defined ``local realism'' 
 in the following way. 
 ``Realism'' is sufficient to state 
 that if one can predict with 
 certainty the result of a measurement of a physical quantity 
 at $A$, without disturbing the 
 system $A$, then the results of the measurement were predetermined and 
 one has an 
 ``element of reality'' corresponding to this 
 physical quantity. The element of reality is a variable which assumes 
 one of a set of values which are the predicted results of the 
 measurement. This value gives the result of the 
 measurement, should it be performed. 
 Locality states that the events at $A$ cannot, 
 instantaneously, disturb 
 in any way a second system at $B$ spatially separated from $A$. Taken together   
 ``local realism'' is sufficient to 
 imply that, if one can predict the result of a measurement of a 
 physical quantity at $A$, by 
 making a simultaneous measurement at $B$, then the result of the 
 measurement at $A$ is described by an element of reality.

Macroscopic local realism $^{\cite{9}}$ is defined 
   as a premise stating the following. If one can predict 
   the result of a measurement at $A$ by performing 
 a simultaneous measurement on a spatially separated system $B$, then the 
 result of the measurement at $A$ is predetermined but 
 described by an element of reality 
 which has an indeterminacy in each of its possible  values, so 
 that only values macroscopically 
 different to those predicted are excluded.
 
 Macroscopic local realism 
 is based on a ``macroscopic locality'', 
 which states that measurements at a 
location $B$ cannot instantaneously induce macroscopic 
changes (for example the dead to alive state of a cat, or a change 
between macroscopically 
different photon numbers) in a second system $A$ spatially 
separated from $B$. 
 Macroscopic local realism also incorporates a 
 ``macroscopic realism'', since it implies elements of reality with (up 
 to) a macroscopic indeterminacy. 
 Suppose 
our ``Schrodinger's cat $^{\cite{5}}$'' is correlated with a second 
spatially separated system, for 
example a gun used to kill the cat.  The strength of 
macroscopic local realism is understood when one realises that its 
rejection in this example means we cannot think of the cat as being 
either dead or alive, even though we can predict the dead or alive 
result of ``measuring'' the cat, without disturbing the cat, 
by a measuring the correlated spatially-separated second system.

   In this paper we present a quantum state which violates a 
   Bell inequality even for coarse measurements with macroscopic 
   uncertainties (in absolute terms), and 
    show how this implies a failure of macroscopic local 
   realism as we define it. 
    Our proposed experiment 
    is depicted in Figure 1a, where ($I_{0}$ is a modified Bessel 
    function and $r_{0}=1.1$)  
  \begin{equation}
	|\psi\rangle = [I_{0}(2r_{0}^{2})]^{-1/2}	|\alpha>_{a_{+}} |\beta>_{b_{+}}
	\left(\sum_{n=0}^{\infty}
	\frac{(r_{0}^{2})^{n}}{n!}|{n}>_{a_{-}} |{n}>_{b_{-}}\right). 
\end{equation}
 The  $\hat a_{\pm}$ and $\hat b_{\pm}$ are boson operators for four  
 outgoing fields. Fields $\hat a_{+}$ and $\hat b_{+}$ are in coherent states  
$|\alpha>_{a_{+}¥}$ and  $|\beta>_{b_{+}¥}$ respectively, and we 
allow $\alpha$, $\beta$ to be real and large. $|{n}>_{k}$ is a Fock 
state for field $k$. The fields $\hat 
a_-$ 
and $\hat b_-$ are microscopic and are generated 
in a pair-coherent state $^{\cite{10}}$. Such 
states are the two-mode equivalent of the 
recently realised $^{\cite{6}}$ ``even'' and ``odd'' coherent 
superposition states 
($(1 \pm exp(-2|\alpha|^2))^{-1/2}¥(|\alpha> \pm |-\alpha>)/\sqrt{2}$) and 
could potentially be generated using 
nondegenerate parametric 
oscillation in a limit where one-photon losses are negligible. 
(The coherent states for $\hat a_{+}¥$ and $b_{+}¥$ would be derived from 
the laser pump for the oscillator.) 
We point out later other choices of $|\psi\rangle$ 
possible.
    
The fields  $\hat a_{\pm}$ are mixed using phase shifts and  
beam splitters to give two new output fields 
$\hat a_{-}^{'}=(\hat a_{-}-\hat a_{+})/\sqrt{2}$ and 
    $\hat a_{+}^{'}=i(\hat a_{-}+\hat a_{+})/\sqrt{2}$
    at the location $A$. Similarly the fields $\hat b_{\pm}$ are mixed 
    to give outputs $\hat b_{\pm}^{'}¥$ at location $B$, spatially 
    separated from $A$. The mixing is incorporated into the experiment 
    simply to provide the nice feature that both 
    fields, $\hat a_{\pm}^{'}¥$ say at $A$, incident on 
    the measuring apparatus are macroscopic.    
      We measure simultaneously at $A$ and $B$ the Schwinger spin 
      operators 
          $\hat S_{\theta}^A= (\hat c_+^\dagger \hat c_+-\hat c_-^\dagger \hat c_-)/2$ 
    and $\hat S_{\phi}^B= (\hat d_+^\dagger \hat d_+-\hat d_-^\dagger \hat 
    d_-)/2$. The measurements are made through the 
transformations (achieved with polarisers or beam 
    splitters with a variable transmission) 
    $c_{+}=\hat a_{+}^{'}\cos{\theta/2}+ \hat a_{-}^{'}\sin{\theta/2}$ 
and $c_{-}=\hat a_{+}^{'}\sin{\theta/2}- \hat a_{-}^{'}\cos{\theta/2}$, at $A$, 
and $d_{+}=\hat b_{+}^{'}\cos{\phi/2}+ \hat b_{-}^{'}\sin{\phi/2}$ 
and $d_{-}=\hat b_{+}^{'}\sin{\phi/2}- \hat b_{-}^{'}\cos{\phi/2}$, at $B$, 
 followed 
by photodetection.

In Figure 1b we demonstrate how the measurement $\hat S_{\theta}^A$ 
can also be 
performed directly from $\hat a_{\pm}¥$ by introducing a relative phase shift 
$\theta$ and mixing with a $50/50$ beam splitter 
to produce $\hat c_{\pm}^{'}=\left( 
\hat a_{+} \pm \hat a_{-} \exp (-i\theta) \right)/\sqrt{2}$, followed by 
photodetection to give $\hat S_{\theta}^A= (\hat c_+^{'\dagger} 
\hat c_+^{'}¥-\hat c_-^{'\dagger} \hat c_-^{'}¥)/2$. 
 
 Our test of macroscopic local realism requires noisy measurements. 
    The result for the photon number differences $
    \hat n_{\theta}^{A}=2\hat S_{\theta}^A=
    \hat c_+^\dagger \hat c_+-\hat c_-^\dagger \hat 
  c_-$ and  
    $\hat n_{\phi}^{B}=2\hat S_{\phi}^B=
    \hat d_+^\dagger \hat d_+-\hat d_-^\dagger \hat 
    d_-$ is of the form $n+noise$, where $n$ is the result of the measurement
    in the absence of the noise. 
    We introduce noise distribution functions at 
  each of $A$ and $B$, and define probabilities such as  
  $P^A(noise\geq x)$, that the $noise$ at $A$ 
   is greater than or equal to the 
  value $x$. A probability $P^B(noise\geq x)$ is defined similarly. 
  Later we allow
 $noise$ to be a random noise term with a gaussian distribution of standard 
 deviation $\sigma$. Photon number measurements for macroscopic fields 
are performed with photodiode 
detectors, which already 
introduce a limited resolution because of 
detection inefficiencies.

    The results of measurements are classified as $+1$ if the photon number 
    difference result  
  is positive or zero, and $-1$ otherwise. 
  We determine the 
following probability distributions: $P_{+}^{A}(\theta)$ 
for obtaining $+$ at $A$; $P_{+}^{B}(\phi)$ for obtaining $+$ at $B$; and 
$P_{++}^{AB}(\theta,\phi)$ the joint probability of obtaining $+$ at 
both $A$ and $B$.

 As a first step we define the  
 probability $P_{ij}^{0,AB}(\theta,\phi)$ for 
 obtaining results $i$ and 
 $j$ respectively upon joint measurement of 
 $\hat n_{\theta}^A$ at $A$, and $\hat n_{\phi}^B$ at $B$, 
  in the absence of the applied noise $\sigma$. With noise present 
  at the detectors, the measured 
 probabilities $P_{++}^{AB}(\theta,\phi)$ become  
   \begin{equation}
   P_{++}^{AB}(\theta,\phi) = \sum_{i,j=-\infty}^{\infty}
   P_{ij}^{0,AB}(\theta,\phi) P^A(noise\geq -i) P^B(noise\geq -j)
\end{equation}
  Before presenting the quantum prediction for these probabilities, we 
  examine the prediction given by macroscopic local realism.
  
   Local realism as originally defined by 
  Einstein-Podolsky-Rosen, Bell and 
  Clauser-Horne $^{\cite{1}}$ implies the following well known expression.  
    \begin{equation}
   P_{ij}^{0,AB}(\theta,\phi) = \int \rho(\lambda) \quad p_{i}^A(\theta, \lambda ) 
   p_{j}^B(\phi, \lambda )\quad d\lambda  
	\label{8}
\end{equation}
Local realism implies an underlying set of elements of reality, or 
hidden variables $\lambda$ (with probability distribution $\rho(\lambda)$),  
not specified by quantum theory. The element of reality is a variable 
which assumes one of a set of values
which are the predicted results of the measurement, $\hat n_{\theta}^{A}$ 
say. For our experiment, a precise prediction of $\hat n_{\theta}^{A}$ 
is not possible given a measurement at $B$, for any choice $\phi$ at 
$B$. The elements of reality then do not take on definite values and 
local realism is only sufficient to imply 
a probability $p_{i}^A(\theta, \lambda )$ for 
 the result $i$ of the measurement $\hat n_{\theta}^{A}$, for a given $\lambda$. 
  The independence of  $p_i^A (\theta, \lambda)$ 
  on $\phi$ is based on the locality assumption. 
   
  Now we consider the prediction given by macroscopic local realism. 
  With macroscopic local realism the locality condition is relaxed, but only up to the level of 
  $M$ photons, where $M$ is not macroscopic, by maintaining 
  that the measurement at $B$ cannot instantaneously 
  change the result at $A$ by an amount exceeding $M$ photons.
 The elements of reality deduced using macroscopic local realism can give 
  predictions for the results of measurement  
  which are microscopically (or mesoscopically) different, but not 
  macroscopically different, to those predicted from the 
  elements of reality deduced using local realism. Where our predicted result 
  at $A$ is 
  $i^{'}¥$ using local realism, macroscopic local realism 
  allows the result to be $i=i^{'}¥+m_{A}¥$ where $m_{A}¥$ can be any 
  number not macroscopic. Importantly, while $i^{'}¥$ is not dependent on 
  the choice $\phi$ at $B$, the nonmacroscopic value $m_{A}¥$ can be. 
   Where local realism specifies a (local) probability distribution 
    $p_{i'}^{A} (\theta, \lambda )$ for obtaining 
   $i'$ photons at $A$, the prediction is only correct to within $\pm M$ 
   photons. The actual result at $A$ is determined by a further nonlocal 
   perturbation term $p_{m_A}^{A} (i^{'},{\theta,\phi},\lambda)$, which gives 
   the probability of a further change of $m_A$ photons. 
The macroscopic local realism assumption then is that the conditional probability  
   $p_{i}^A(\theta, \lambda )$ in equation (3) is expressible as the 
   convolution (and similarly for  $p_i^B (\phi, \lambda)$): 
       \begin{equation}
     p_{i}^A({\theta,\phi}, \lambda )= \sum_{m_A=-M}^{+M}
      p_{m_A}^{A} (i^{'},{\theta,\phi},\lambda) 
      p_{i'=i-m_A}^{A} (\theta, \lambda ). 
	\label{9}
\end{equation}
The original local probability $p_{i'}^{A} (\theta, \lambda )$ 
can be convolved with a microscopic 
nonlocal probability function $p_{m_A}^{A} (i^{'},{\theta,\phi},\lambda)$, 
the only restriction being that the nonlocal distribution does not 
provide macroscopic perturbations, so that the 
	probability of getting a nonlocal change outside the range $m_A=-M,...,
	+M$ is zero. 
	Equivalently we must have (and similarly for terms with $B$)  
\begin{equation}	
	 \sum_{m_A=-M}^{M}
       p_{m_A}^{A} (i^{'},{\theta,\phi},\lambda)  = 1.   
  \end{equation}   
We substitute the macroscopic locality 
assumption (4) into the hidden variable prediction (3) to obtain the 
prediction for the measured probabilities (2).
    \begin{eqnarray}
   P_{++}^{AB}(\theta,\phi) &=& 
   \sum_{i,j=-\infty}^{\infty} \int \rho(\lambda)\biggl[ \sum_{m_A=-M}^{M}
       p_{m_A}^{A} (i^{'},{\theta,\phi},\lambda) 
      p_{i'=i-m_A}^{A} (\theta, \lambda )   \biggr.\nonumber \\
&\times&  \biggl.\sum_{m_B=-M}^{M} p_{m_B}^{B} (j^{'},{\phi,\theta},\lambda) 
      p_{j'=j-m_B}^{B} (\phi, \lambda )\biggr] d\lambda \quad 
       P^A(noise\geq -i) P^B(noise\geq -j) 
	\label{11}
\end{eqnarray}
Recalling $i=i'+m_A$ and $j=j'+m_B$ we change the $i$, $j$ summation to 
one over $i'$, $j'$ to get
  \begin{eqnarray}
   P_{++}^{AB}(\theta,\phi) &=& 
   \sum_{i',j'=-\infty}^{\infty}\int \rho(\lambda)p_{i'}^{A} (\theta,\lambda)
    \biggl[\sum_{m_A=-M}^{M}
        p_{m_A}^{A} (i^{'},{\theta,\phi},\lambda) 
       P^A(noise\geq -(i'+m_A)) \biggr]  \biggr. \nonumber \\
 &\times&   p_{j'}^{B} (\phi, \lambda )\biggl[\sum_{m_B=-M}^{M}
\biggl.     p_{m_B}^{B} (j^{'},{\phi,\theta},\lambda) 
     P^B(noise\geq -(j'+m_B)) \biggr] d\lambda  
      \nonumber \\
 	\label{12}
\end{eqnarray}
We assume that 
 the noise function $noise$ is slowly varying over the 
microscopic (or mesoscopic) range $-m_{A},..+m_{A}$ for 
which nonlocal perturbations are 
possible according to macroscopic local realism (and similarly at $B$):   
 \begin{equation}
\sum_{m_A=-M}^{M}
        p_{m_A}^{A} (i^{'},{\theta,\phi},\lambda) 
       P^A(noise\geq -(i'+m_A)) \approx 
        P^A(noise\geq -i') \sum_{m_A=-M}^{M}
        p_{m_A}^{A} (i^{'},{\theta,\phi},\lambda).
   \end{equation}      
This is only valid 
if $\sigma$ is macroscopic.
Using (5), one simplifies to get the final form $
   P_{++}^{AB}(\theta,\phi)=\sum_{i',j'}\ \int \rho(\lambda)
     p_{i'}^{A} (\theta, \lambda )  p_{j'}^{B} (\phi, \lambda ) d\lambda 
    \times P^A(noise\geq -i') P^B(noise\geq -j')$. 
  This prediction of the hidden variable 
	theory is now given in a (local) form like that of (3), from which Bell-
	Clauser-Horne  
	inequalities $^{\cite{1}}$ follow, for example:
	  \begin{equation}
 S={{P_{++}^{AB}(\theta,\phi)-P_{++}^{AB}(\theta,\phi')+P_{++}^{AB}(\theta',\phi)
 +P_{++}^{AB}(\theta',\phi')}\over{P_{+}^{A}(\theta')+P_{+}^{B}(\phi)}} \leq 1.
	\label{eqnbell}
\end{equation} 

	The noise terms which add a macroscopic 
	uncertainty to the photon number result alter the premises 
	needed to derive the Bell inequality. With $\sigma$ macroscopic we
	 need only assume 
	macroscopic local realism to derive the Bell 
	inequality (9).

The quantum prediction for state (1) is shown in Figure 2.
Violations of the Bell inequality (9) in the absence 
 of $noise$ are shown in curve (a).
Violations are still  
 possible (curve (b)) in the presence of increasingly 
 larger absolute noise $\sigma$, simply by 
 increasing $\alpha=\beta$.
This violation of the Bell inequality (9) with macroscopic noise 
$\sigma$ implies the failure of macroscopic local 
	realism.

The asymptotic behavior in the large $\alpha$,$\beta$ limit is 
crucial in determining a violation of macroscopic local realism, and 
is best understood by replacing the boson 
operators $\hat a_+$ and $\hat b_+$ with classical amplitudes $\alpha$ and 
$\beta$ 
respectively. We see that then 
$\hat S_{\theta}^A=\alpha \hat X_{\theta}^A/2$ and  
$\hat S_{\phi}^B=\beta \hat X_{\phi}^{B}/2$, 
where  $\hat X_{\theta}^A=\hat a_-exp(-i\theta)+\hat a_-^\dagger exp(i\theta)$ 
and $\hat X_{\phi}^B=\hat b_-exp(-i\phi)+\hat b_-^\dagger exp(i\phi)$ are 
the quadrature phase amplitudes of fields $\hat a_{-}¥$ and $\hat b_{-}¥$. 
In fact Figure 1b with $\alpha$,$\beta$ large 
shows the experimental set-up for balanced 
homodyne detection of the quadrature phase amplitudes $\hat X_{\theta}^A$ 
and $\hat X_{\phi}^B$, of fields $\hat a_{-}¥$ and $\hat b_{-}¥$. 
Homodyne detection has been used experimentally to detect ''squeezed'' 
fields $^{\cite{11}}$, where the fluctuation in $\hat X_{\theta}^A$ is reduced below the 
standard quantum limit. 

 Violations of Bell inequalities for    
measurements $\hat X_{\theta}^A$, $\hat X_{\phi}^B$ on state (1) 
have recently been predicted $^{\cite{12}}$, confirming Figure 2(a) 
in the large $\alpha$ limit. These violations vanish $^{\cite{13}}$ when 
gaussian noise of standard deviation $\sigma_{0}\geq 0.26$ is added to 
the measurements $\hat X_{\theta}^A, \hat X_{\phi}^B$.  
With $\alpha$ large, this 
corresponds to a noise value of $\alpha \sigma_{0}¥$ in the photon 
number measurement $2\hat S_{\theta}^A$, confirming Figure 2(b). 
 In fact, since there is always a finite 
cutoff $\sigma_{0}¥$, any state $|\psi\rangle$ which shows a failure of local realism for 
measurements  $\hat X_{\theta}^A$ 
and $\hat X_{\phi}^B$ on fields $\hat a_{-}¥$ and $\hat b_{-}¥$ will also show a 
violation of macroscopic local realism, provided $\alpha$,$\beta$ are 
large. Other such states have been recently 
predicted $^{\cite{14}}$, increasing the scope for a practical violation of 
macroscopic local realism.

\begin{figure}%
\includegraphics[scale=.85]{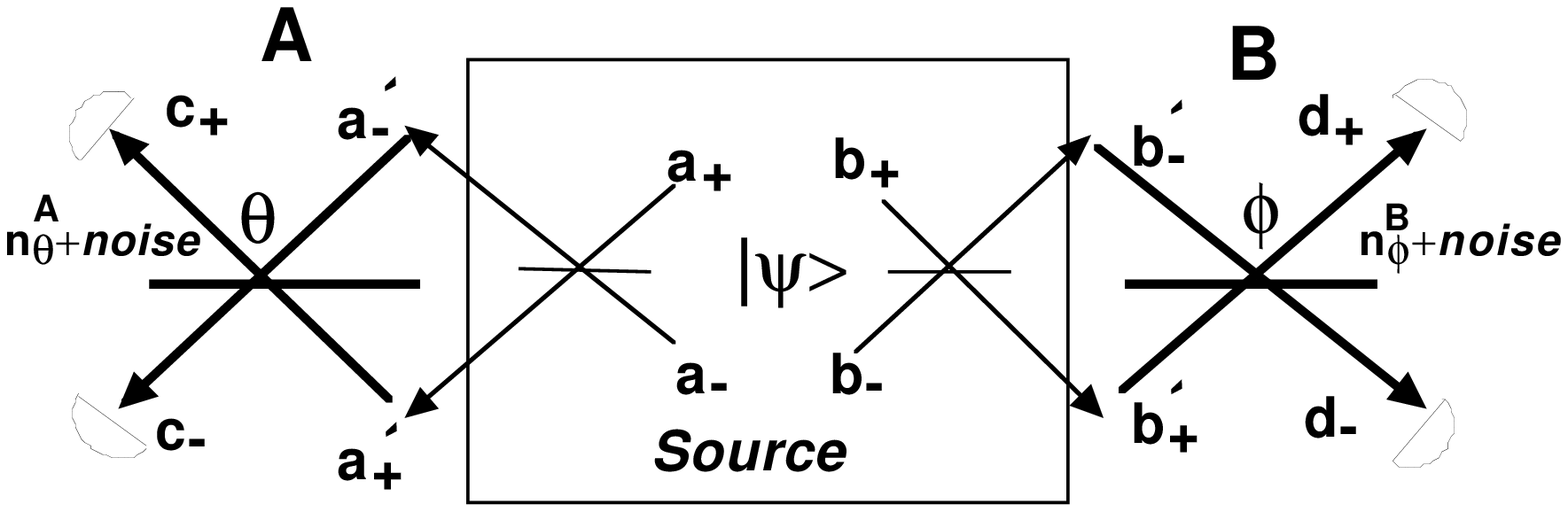}
\includegraphics[scale=.85]{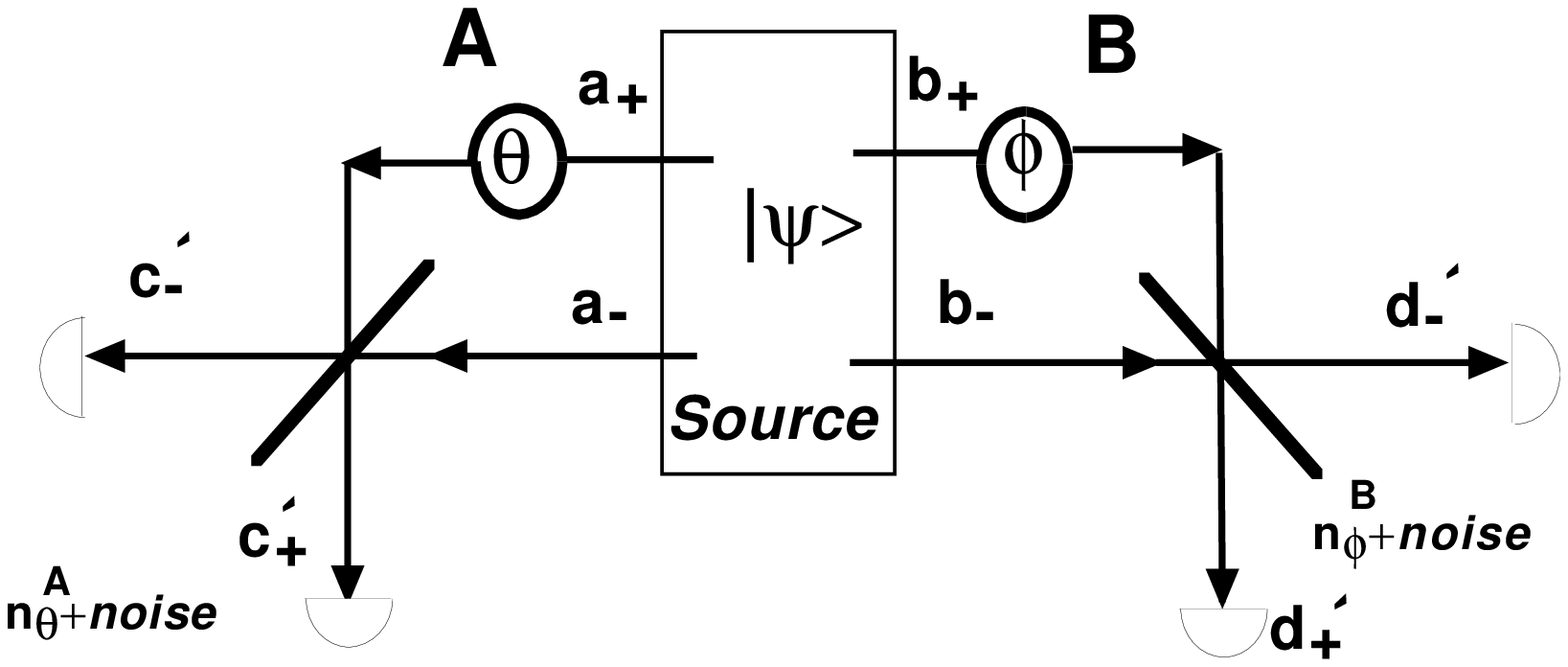} 
\caption[Schematic representation of a test of macroscopic local 
realism.]
{Our proposed test of macroscopic local 
realism. 
 (a) For large $\alpha,\beta$ macroscopic fields 
 $a_{\pm}^{'}$ ($b_{\pm}^{'}$) are incident on each measuring apparatus.
  (b) This measurement scheme for $\alpha,\beta$ large corresponds to 
  balanced homodyne 
 detection of the quadrature phase amplitudes $\hat X_{\theta}^A$ 
and $\hat X_{\phi}^B$.
 }%
\end{figure}

\begin{figure}
\includegraphics{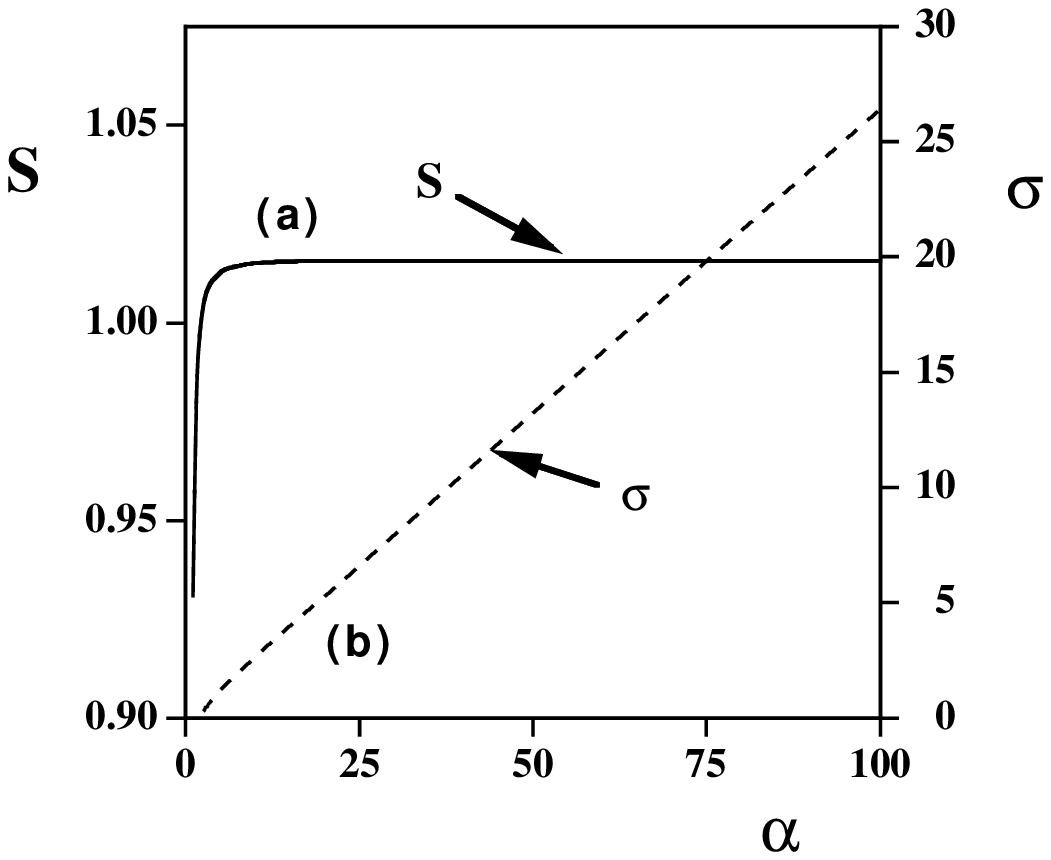}
\caption[Violations of the Bell's inequality found 
for the quantum state (1).]
{(a) $S$ versus $\alpha=\beta$, for 
$\theta=0,\phi=-\pi/4,\theta'=\pi/2,\phi'=-3\pi/4$ for the 
quantum state (1) with no noise present.
(b) Maximum noise $\sigma$ still giving a violation of the Bell 
inequality (9), versus $\alpha$.
}%
\end{figure}

\end{document}